\newcommand{\lya}{Ly$\alpha$}

\documentstyle[12pt,blois,psfig]{article}

\begin{document}

\heading{GALAXIES AT REDSHIFTS $z > 5$}

\author{Kenneth M. Lanzetta$^{1}$, H.-W. Chen$^{1}$, S. Pascarelle$^{1}$, N.
Yahata$^{1}$}{$^{1}$ State University of New York at Stony Brook, Stony Brook,
U.S.A.}

\begin{bloisabstract}

  Here we describe our attempts to establish statistically complete samples of
very high redshift galaxies by obtaining photometric redshifts of galaxies in
Medium Deep Survey (MDS) fields and photometric and spectroscopic redshifts of
galaxies in very deep STIS slitless spectroscopy fields.  On the basis of this
analysis, we have identified galaxies of redshift $z = 4.92$ in an MDS field
and of redshift $z = 6.68$ in a very deep STIS field.

\end{bloisabstract}

\section{Introduction}

  Our photometric redshift technique applied to observations of the Hubble Deep
Field (HDF) has identified galaxies at redshifts up to and beyond $z = 6$
(\cite{lyf96}, \cite{fly98}).  The recent confirmation of two of these
measurements at redshifts $z > 5$ (\cite{weymann98}, \cite{spinrad98}) has
established that broad-band redshift measurement techniques can accurately and
reliably identify high-redshift galaxies---not only in a general sense,
distinguishing high- from low-redshift galaxies, but also in a specific sense,
establishing redshifts to within relative errors of typically $\Delta z / (1+z)
< 15\%$.

  Building on the success of the broad-band redshift measurement techniques, we
have sought to establish statistically complete samples of very high redshift
galaxies by means of four distinct methods applied to four distinct
collections of observations:  (1)  photometric redshifts of galaxies in the HDF,
(2) photometric redshifts of galaxies in Medium Deep Survey (MDS) fields, (3)
photometric and spectroscopic redshifts of galaxies in very deep STIS slitless
spectroscopy fields, and (4) photometric redshifts of infrared detected (and
optical non-detected) galaxies in the HDF.  In principle, the
optical-wavelength observations are sensitive to galaxies at redshifts as large
as $z \approx 7$, while the infrared-wavelength observations are sensitive to
galaxies at redshifts as large as $z \approx 17$.

  Here we describe initial results of the second and third of these programs,
namely of photometric redshifts of galaxies in MDS fields and photometric and
spectroscopic redshifts of galaxies in very deep STIS slitless spectroscopy
fields.

\section{Photometric redshifts of galaxies in MDS fields}

  First, we have sought to measure photometric redshifts of galaxies in MDS
fields.  The MDS (\cite{ratnatunga98}) currently spans more than 450 fields,
which have been observed using HST with WFPC2 and the F814W, F606W, and (in
some cases) the F450W filters.  These observations are, of course, not as deep
as the observations of the HDF.  But what these observations lack in depth,
they make up for in breadth, covering an angular area of $\approx 0.6$ deg$^2$
to a $5 \sigma$ point source limiting magnitude threshold of typically
$AB(8140) \approx 27.0$.  Our previous application of the photometric redshift
technique to the HDF (\cite{lyf96}, \cite{fly98}) incorporated observations
spanning seven photometric bands.  In contrast, our application of the
photometric redshift technique to the MDS fields incorporates observations
spanning only two or three photometric bands, which of course makes photometric
redshift measurement more difficult.  But given sufficiently small photometric
errors, continuum break features indicative of high-redshift galaxies (i.e.\
the Lyman limit and the \lya\ decrement) can be {\em unambiguously}
distinguished from continuum break features indicative of low-redshift galaxies
(i.e.\ the 4000 \AA\ break) because the amplitude of the largest continuum
breaks observed in low-redshift galaxies or main-sequence stars is $\approx 3$
(see \cite{spinrad98} and references therein), whereas the amplitude of the
Lyman continuum break is in principle infinite or close to it.  The key, then,
is to apply the two- or three-band photometric redshift technique only in cases
where the photometric errors are small enough to distinguish the Lyman break
from other continuum features to a high level of significance---i.e.\ to bright
(through the F814W filter) galaxies or to deep images.  We have so far applied
this analysis to a portion of the MDS observations in order to identify
candidate galaxies for confirming spectroscopy.

  Figure 1 shows a spectrum obtained with the KPNO 4 m telescope of a
high-redshift galaxy identified in an MDS field.  The spectrum is characterized
by an emission line at $\lambda = 7200$ \AA, which we interpret as \lya, and
by a continuum break at $\lambda = 7200$ \AA, which we interpret as the \lya\
decrement, in which case the redshift of the galaxy is $z = 4.92$.  Based on
an initial analysis, we expect that $\approx 65$ galaxies of redshift $z > 5$
be identified in the MDS fields.

\begin{figure}
\centerline{\psfig{file=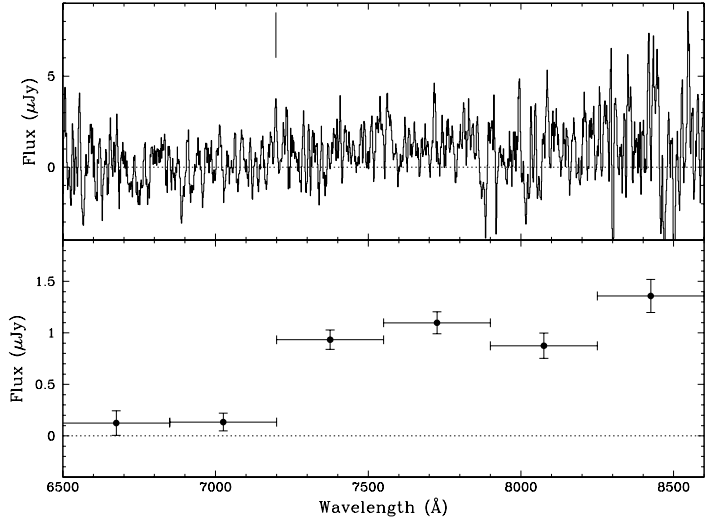,width=4in,angle=-90}}
\caption{Spectrum obtained with the KPNO 4 m telescope of a high-redshift
galaxy identified in an MDS field.  Upper panel shows spectrum (of pixel size 5
\AA\ boxcar smoothed by three pixels), and lower panel shows spectrum cast into
350 \AA\ bins.  Tick mark in upper panel indicates a statistically significant
emission line.  We interpret the emission line at $\lambda = 7200$ \AA\ as
\lya\ and the continuum break at $\lambda = 7200$ \AA\ as the \lya\ decrement,
in which case the redshift of the galaxy is $z = 4.92$.}
\end{figure}

\section{Photometric and spectroscopic redshifts of galaxies in very deep STIS
slitless spectroscopy observations}

  Second, we have sought to measure photometric and spectroscopic redshifts of
galaxies in very deep STIS sliless spectroscopy fields.  At near-infrared
wavelengths, where background sky light is the dominant source of noise, the
Hubble Space Telescope (HST) using the Space Telescope Imaging Spectrograph
(STIS) is more sensitive than the Keck telescope because from space (1) the sky
is fainter and (2) the spatial resolution is higher.  To exploit the unique
sensitivity of STIS at near-infrared wavelengths, the Space Telescope Science
Institute and the STIS instrument team at the Goddard Space Flight Center
initiated the STIS Parallel Survey, in which deep STIS observations are
obtained in parallel with other observations \cite{gardner98}.  From among the
observations so far obtained by the Survey, we selected for analysis very deep
observations acquired in slitless spectroscopy mode, because these observations
are best suited for identifying distant galaxies.

  The difficulty of slitless spectroscopy is that the objects can overlap along
the dispersion direction.  We have overcome this difficulty by developing and
applying a new method of analyzing slitless spectroscopy observations that
makes optimal use of the direct and dispersed images that are recorded as part
of a normal sequence of observation.  Specifically, we use the direct image to
determine not only the exact locations but also the exact two-dimensional
spatial profiles of the spectra on the dispersed image.  These spatial profiles
are crucial because they provide the ``weights'' needed to optimally extract
the spectra and the models needed to deblend the overlapping spectra and
determine the background sky level.  Our analysis proceeds in three steps:
First, we identify objects in the direct image, using standard source
extraction programs.  Next, we model each pixel of the dispersed image as a
linear sum of contributions from relevant portions of all overlapping
neighboring objects and background sky.  Finally, we minimize $\chi^2$ between
the model and the data with respect to the model parameters to form optimal
one-dimensional spectra.

  We measure redshifts from the optimal one-dimensional spectra by means of
{\em both} broad-band continuum features and narrow-band emission and absorption
features.  Specifically, we measure photometric redshifts using a variation of
the photometric redshift technique described previously by \cite{lyf96}, and we
seek to verify the photometric redshifts by identifying confirming narrow
emission and absorption features.  At low redshifts, the most prominent
broad-band feature is the 4000 \AA\ break and the most prominent narrow-band
features are the [O II] emission line and the Ca II H and K absorption lines,
while at high redshifts, the most prominent broad-band feature is the \lya\
decrement and the most prominent narrow-band feature is the \lya\ emission
line.

  Figure 2 shows examples of photometric and spectroscopic redshifts of
galaxies in very deep STIS slitless spectroscopy fields.  The spectrum of the
highest-redshift galaxy is characterized by an emission line at $\lambda =
7200$ \AA, which we interpret as \lya, and by a continuum break at $\lambda =
7200$ \AA, which we interpret as the \lya\ decrement, in which case the
redshift of the galaxy is $z = 6.68$.  Based on an initial analysis, we expect
that redshifts of a large fraction of galaxies in the STIS field will be
identified.

\acknowledgements{This research was supported by NASA grant NAGW--4422 and NSF
grant AST--9624216.}

\begin{figure}
\centerline{\psfig{file=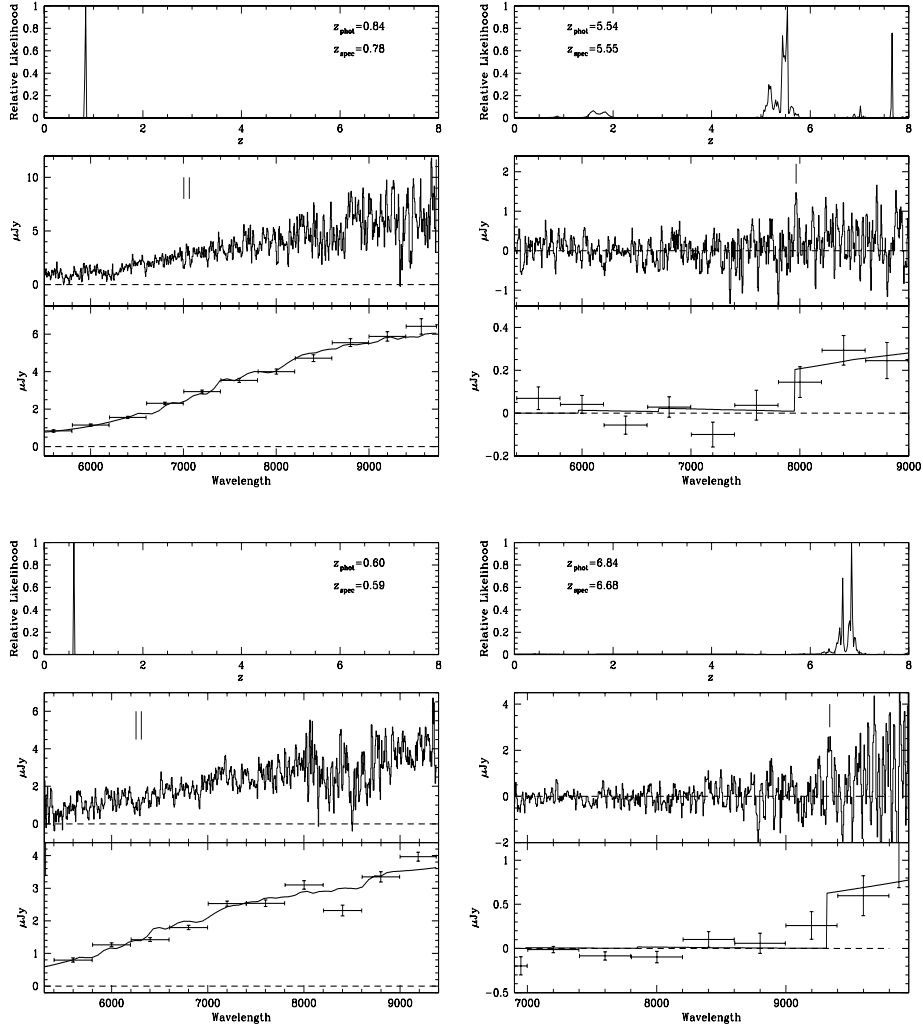,width=5in}}
\caption{Examples of photometric and spectroscopic redshifts of galaxies in
very deep STIS slitless spectroscopy fields.  For each example, top panel shows
redshift likelihood function, middle panel shows spectrum (of pixel size 5 \AA\
boxcar smoothed by three pixels), and bottom panel shows spectrum cast into
400 \AA\ bins together with best-fit spectrophotometric template spectrum.
Tick marks indicate statistically significant emission or absorption lines.}

\end{figure}

\vfill
\end{document}